\documentclass{www2012-comp-accepted}

\usepackage{subfigure}
\usepackage{graphicx}
\usepackage{multirow}
\usepackage{comment}
\usepackage{amssymb}
\usepackage{multirow}
\usepackage{color}
\usepackage{rotating}
\usepackage{amsmath}
\usepackage{afterpage}
\usepackage{multirow}

\usepackage{graphicx,epsfig}
\usepackage{graphics,dcolumn,bm,epic,eepic,float}
\usepackage{amssymb,amsmath,multirow,rotate,color}
\usepackage{subfigure}
\usepackage{epstopdf}

\usepackage{etex}
\reserveinserts{100}
\usepackage{morefloats}
\usepackage{paralist}
\usepackage{hyperref}
\hypersetup{colorlinks=true,urlcolor=blue,citecolor=blue,linkcolor=blue}
\usepackage{xspace}
\usepackage{tabularx}
\usepackage{booktabs}

\newcommand{\mat}[1]{\boldsymbol{#1}}
\renewcommand{\vec}[1]{\boldsymbol{\mathrm{#1}}}

\providecommand{\mA}{\ensuremath{\mat{A}}}

\providecommand{\mE}{\ensuremath{\mat{E}}}

\providecommand{\mG}{\ensuremath{\mat{G}}}

\providecommand{\mM}{\ensuremath{\mat{M}}}

\providecommand{\ones}{\vec{e}}

\providecommand{\vx}{\ensuremath{\vec{x}}}

\newcommand\TT{\rule{0pt}{2.3ex}}
\newcommand\BB{\rule[-1.0ex]{0pt}{0pt}}

\begin{document}
\clubpenalty=10000 
\widowpenalty = 10000

\title{{\ttlit Role-Dynamics:} Fast Mining of Large Dynamic Networks}

\numberofauthors{2}
\author{
\alignauthor
Ryan Rossi \\
\vspace{1.5mm}
Jennifer Neville\\
\vspace{3mm}
       \affaddr{Purdue University}\\
       \email{\{rrossi, neville\}@purdue.edu}
\alignauthor
Brian Gallagher \\
\vspace{1.5mm}
Keith Henderson\\
\vspace{3mm}
       \affaddr{Lawrence Livermore Lab}\\
       \email{\{bgallagher, keith\}@llnl.gov}
}

\maketitle
\begin{abstract} 
To understand the structural dynamics of a large-scale social, biological or technological network, it may be useful to discover behavioral roles representing the main connectivity patterns present over time. 
In this paper, we propose a scalable non-parametric approach to automatically learn the \textit{structural dynamics} of the network and individual nodes. 
Roles may represent structural or behavioral patterns such as the center of a star, peripheral nodes, or bridge nodes that connect different communities.
Our novel approach learns the appropriate structural ``role'' dynamics for any arbitrary network and tracks the changes over time.
In particular, we uncover the specific global network dynamics and the local node dynamics of a technological, communication, and social network.
We identify interesting node and network patterns such as stationary and non-stationary roles, spikes/steps in role-memberships (perhaps indicating anomalies), increasing/decreasing role trends, among many others.
Our results indicate that the nodes in each of these networks have distinct connectivity patterns that are non-stationary and evolve considerably over time. 
Overall, the experiments demonstrate the effectiveness of our approach for fast mining and tracking of the dynamics in large networks.
Furthermore, the dynamic structural representation provides a basis for building more sophisticated models and tools that are fast for exploring large dynamic networks.
\end{abstract} 



\section{Introduction}

Many social, biological, and technological networks contain dynamics that are important to model.
The links, nodes, and attributes of these dynamical systems change considerably as time progresses.
Naturally, these dynamic networks induce arbitrary patterns of connectivity that are challenging to identify in an automated adaptive fashion.
To complicate matters further, the patterns observed in these dynamic networks are not necessarily stationary and may change considerably.
In addition, these networks are usually large and a significant amount of data is continuously collected.
All of the above issues warrant a fast completely automatic approach for identifying and tracking arbitrary patterns in large dynamic networks.

We address the problem of representing, tracking, and analyzing the structural dynamics of these large networks in a fast and completely-automatic manner. 
Our approach captures arbitrary patterns of connectivity, has no parameters, and is fast for large networks (linear in the number of edges).
At the heart of our framework is the representation of the structural dynamics, which can be used to build a variety of sophisticated analysis tools.
We use the structural dynamics framework for exploring the evolution of the network and individual nodes. 
Our approach lends itself to visualizations that clearly show how the behavior of individual nodes and the network as a whole change over time.

Consider a large dynamic (or streaming) network, how can we automatically learn the temporal structural behaviors of individual nodes and identify unusual activities or patterns?
For instance, in an IP-to-IP network, we may want to learn the ``behavioral roles'' of individual hosts and monitor the changes over time. 
This would allow us to characterize the dynamic behaviors of individual hosts and also detect when a machine or host becomes compromised or begins having unusual behavior with respect to the global network dynamics (as well the dynamics of that local host). 

For capturing the behavior of large time-evolving networks, we propose a \textit{structural dynamics framework} that essentially (1) extracts node features from a sequence of graphs over time, (2) discovers roles from the sequence of node-by-feature matrices over time, (3) tracks the node memberships over time, and (4) captures the temporal dependencies of the nodes and roles over time. 
Our novel algorithm tracks the network dynamics and the behavioral roles of individual nodes over time. 
Behavioral roles or more precisely structural patterns are defined as a combination of similar structural features that were learned from the initial network.
Since similar structural features are combined into a single role, then each role represents a different structural pattern (or connectivity pattern).
More specifically, the roles represent similar network features that were recursively extracted automatically. 
Therefore, if two nodes share a common role at a given timestep, then these two nodes are structurally similar.

Our novel approach provides a basis for analyzing the local node dynamics and the global network dynamics as a whole.
Network dynamics refers to the structural patterns present in the entire network as it evolves over time. 
We posit that the importance of the learned structural patterns fluctuate and eventually change entirely. 
For example, the structural behavioral dynamics present in the initial Twitter social network in 2006 are most likely different from the structural dynamics observed today. 
This difference could be due to changes in their privacy policy, or the addition of features for twitter users or applications for mobile devices, etc. 
The frequency of the fluctuations and changes depends entirely on the dynamical system (e.g., social or technological network).
In contrast, node dynamics refers to the structural patterns of individual nodes over time. We posit that the nodes structural behavioral dynamics are non-stationary, that is they change or fluctuate over time. For instance, the structure induced by emails for a given user may change during the work hours. Perhaps this user serves as a coordinator at work and therefore during the day their email activity represents structural behaviors such as the center of a star (node with large number of incoming or outgoing edges) or a bridge that connects multiple communities (or departments in this case).

\vspace{2mm}
\noindent The main contributions of our approach are as follows:
\vspace{1mm}
\begin{compactenum}[1.\leftmargin=0em]
\item \textbf{Flexible.} The fast analytical framework for exploring dynamics can serve as a foundation for many other applications and tools.
\item \textbf{Non-parametric and data-driven.} The important structural behavior of the given \textit{temporal network} is represented without having to specify any features/patterns, making it applicable for exploring any type of network, and perhaps more importantly, making it a suitable candidate for real-time anomaly detection.
\item \textbf{Efficient.} The algorithm is linear in the number of edges and thus practical for large real-world networks.
\item \textbf{Automatic.} The algorithm doesn't require user-defined parameters.
\item \textbf{Interpretable.} 
The roles can be interpreted with respect to simple traditional measures. 
The approach lends itself to visualizations that clearly show how the behavior of individual nodes and the network as a whole change over time.
Interesting dynamic connectivity patterns are found in a social, technological, and communication network. 
The patterns are shown to be meaningful and agree with human intuition.
\end{compactenum}
\vspace{2mm}

There has been an increased interest in mining, predicting, and exploiting the temporal nature of datasets~\cite{götz2009modeling,leskovec2005graphs,leskovec2007dynamics,papadimitriou2005streaming,sun2007graphscope,dunlavy2011temporal,abello2010detecting,lin2009analyzing,tang2008community,greene2010tracking}. Most of this work has focused on mining or modeling one aspect of temporal data such as the importance of nodes over time or attributes. In contrast to this work, we propose a scalable non-parametric exploratory analysis method capable of discovering structural dynamic patterns and trends automatically in large time-evolving networks. 
Therefore, our method is appropriate for mining social networks, communication networks, biological networks, among many others. 

\vspace{2mm}\noindent
In addition to being applicable for a wide-range of domains, the structural dynamics framework can be used for a variety of applications, such as:
\vspace{1mm}
\begin{compactenum}[$\circ$ \leftmargin=0em]
\item\textit{Dynamic Network Analysis.} Our approach captures dynamic behavioral patterns of nodes (e.g., a peripheral node becomes an articulation point connecting two communities) and the global network dynamics.
\item\textit{Anomaly Detection.} Identification of nodes or time periods with unusual structural behavior with respect to the global network dynamics.
\item\textit{Sampling.} The representation can be used to sample nodes from each of the learned network behaviors (principle of diversity) and adjust the sample dynamically as the behavioral roles change. 
This sampling strategy could be utilized for active learning on large networks. 
\item \textit{Graph Similarity.} Given two sequences of graphs (or graphs from different generators), we can measure the divergence between the learned features and behavioral patterns. As a simple example, if we learn 10 features for the Internet AS topology, and 100 from a topology of the same size from a generator, then clearly the topology from the generator has more complex connectivity than the true Internet AS.
\item \textit{Generalizations.} We may learn the structural roles on a single social network (Facebook), and use these roles to analyze the dynamics of another social network (Google+). This indicates whether these two social networks are governed by similar social processes (e.g., homophily).
\item \textit{Compact Representation.} In the case of very large networks, the temporal network representation provides a compact and reasonable approximation of the most important graph properties and behavioral patterns. The representation serves as a foundation for building additional large-scale tools and models for exploring and visualizing dynamic networks.
\end{compactenum}
\vspace{2mm}

In this paper, we focus on using the structural dynamics for the first application. The others are left for future work. Section~\ref{sec:tbm} provides a formal definition of our structural dynamics framework while Section~\ref{sec:results} reports results using the framework for exploratory analysis. In Section~\ref{sec:related-work} we discuss related work and in Section~\ref{sec:conclusion} we give some concluding remarks and future directions.

\section{Structural Role Dynamics}\label{sec:tbm}
Given a sequence of graphs, the \textit{structural dynamics framework} (or simply \textsc{Role-Dynamics}) automatically learns a set of representative features, extracts these features from each graph, then discovers behavioral roles, and iteratively extracts these roles from the sequence of node by feature matrices over time. The proposed framework is flexible in that any technique that learns a representative set of features (i.e., searches over the space of node and link features) and role discovery technique can be used instead of the chosen one. In this paper, we use ReFeX~\cite{refex2011kdd} and RolX~\cite{rolX2011-TR} since both have been designed implicitly for large graphs.
Next we formally define the components of the structural dynamics framework.

\subsection{Data Model for Temporal Networks}
Networks accumulate a large number of edges and nodes over time. However at any given time, many of these edges and nodes are inactive. Nodes and edges can appear or disappear at any time. If a given node does not contain any active edges at time $t$ then it is effectively not considered. No assumption is made prior about the number of nodes or edges over time. Edges can be weighted or unweighted and be instantaneous or last for some duration. Multiple edges may exist between nodes. Nodes and edges may have attribute data associated with them that could also be temporal. A snapshot graph is defined by the nodes and edges active at time $t$. 
In the most general case, we have an ordered sequence of snapshot graphs represented as adjacency matrices $\mA_{t}$ for $t = 1,2,...,t_{max}$ where a nonzero $i, j$ entry records the presence or weight of a link from node $i$ to $j$.

\begin{table}[t!]
\vspace{-2mm}
\caption{Summary of notation. Matrices are bold, upright roman letters, vectors are bold, lowercase roman letters, and scalars are unbolded roman or greek letters. Sets are uppercase calligraphy letters.}
\vspace{1mm}
\label{table:notation}
\begin{center}
\begin{small}
\begin{tabular}{ r p{60mm} }
\noalign{\hrule height 1pt}
\TT \BB \multirow{1}{*}{\textbf{$n$}} & number of nodes in a graph  \\
\TT \BB \multirow{1}{*}{\textbf{$f$}} & number of learned features  \\
\TT \BB \multirow{1}{*}{\textbf{$r$}} & number of learned roles  \\
\TT \BB \multirow{1}{*}{\textbf{$\mathbf{\mathcal{G}}$}} & set of node by role matrices  \\
\TT \BB \multirow{1}{*}{\textbf{$\mathbf{\mathcal{V}}$}} & set of node by feature matrices  \\
\TT \BB \multirow{1}{*}{\textbf{$\mathbf{F}$}} & feature by role matrix  \\
\noalign{\hrule height 1pt}
\TT \BB \multirow{1}{*}{\textbf{$n_t$}} & number of active nodes at time $t$  \\
\TT \BB \multirow{1}{*}{\textbf{$\mathbf{A}_t$}} & adjacency matrix at time $t$  \\
\TT \BB \multirow{1}{*}{\textbf{$\mathbf{V}_t$}} & node by feature matrix extracted at time $t$  \\
\TT \BB \multirow{1}{*}{\textbf{$\mathbf{G}_t$}} & node by role matrix extracted at time $t$  \\
\TT \BB \multirow{1}{*}{\textbf{$\mathcal{L}_t$}} & set of discovered features at time $t$  \\
\TT \BB \multirow{1}{*}{\textbf{$\mathbf{D}_t$}} & role (or node/time) distance matrix  \\
\end{tabular}
\end{small}
\end{center}
\vspace{-6mm}
\end{table}

\subsection{Dynamic Behavioral Representation}
We define our representation for dynamic networks, which includes discovering a set of representative features and extracting ``structural roles'' from this large set of features.

\vspace{1mm}\noindent\textit{Feature Discovery and Extraction.}
We use ReFeX~\cite{refex2011kdd} to discover a representative set of features.
In particular, we start with degree features (in/out, unweighted/weighted, and total) and egonet features. 
The egonet includes the node, its neighbors, and any edges in the induced subgraph on these nodes. 
Egonet features include the number of in/out egonet edges and the total egonet edges as well as weighted versions of these features if the edges are weighted. 
Next, we aggregate the existing features of a node using sum/mean and use them to generate new recursive features. 
After each aggregation step, the algorithm prunes redundant features. 
The aggregation proceeds recursively over the current feature set, until no new features are retained. 

For dynamic networks, we can learn a representative set of features and then extract them for each graph over time.
More formally, given a time-evolving network and any known attributes, we discover a set of features denoted $\mathcal{L}$ at time $t$ and extract a node by feature matrix denoted $V_t$ of size $n_t \times f$ where $n_t$ is the number of active nodes and $f$ is the number of features. 
The features for each network snapshot are extracted resulting in a sequence of node-by-feature matrices, denoted $\mathcal{V} = \{\mathbf{V}_t: t = 1,..., t_{max}\}$. The set of learned graph features for each timestep are shown to be minimal and representative. The graph features capture local, community-level, and global properties of the temporal network (through recursive aggregates).

\vspace{2mm}
\noindent\textit{Structural Roles.}
Using the representative set of graph features, we discover structural roles using Non-negative Matrix Factorization (NMF) with Minimum Description Length (MDL) model selection criterion (see RolEx~\cite{rolX2011-TR}).  More formally, given a nonnegative matrix $\mathbf{V}_t \in \mathbb{R}^{n_t \times f}$ and a positive integer $r < \min(n_t,f)$, find nonnegative matrices $\mathbf{G}_t \in \mathbb{R}^{n_t \times r}$ and $\mathbf{F} \in \mathbb{R}^{r \times f}$ that minimizes the functional,
\[
f(\mathbf{G}_t, \mathbf{F}) = \frac{1}{2}||\mathbf{V}_t - \mathbf{G}_t\mathbf{F}||_{F}^{2}
\]

The number of roles $r$ is automatically selected using MDL. Intuitively, learning more roles, increases model complexity, but decreases the amount of errors. Conversely, learning less roles, decreases model complexity, but increases the amount of errors.
In this way, MDL selects the number of behavioral roles $r$ such that the model complexity (\# of bits) and model errors are balanced. Naturally, the best model minimizes, $\# \; of \; bits + errors$. See~\cite{rolX2011-TR} for more details.

The learned role-by-feature matrix $\mathbf{F} \in \mathbb{R}^{r \times f}$ represents the contribution of each role on the extracted features.
After learning these role definitions $\mathbf{F}$, we iteratively estimate node-by-role memberships for each network snapshot $\mathcal{G} = \{\mathbf{G}_t: t = 1,..., t_{max}\}$ given $\mathbf{F}$ and $\mathcal{V} = \{\mathbf{V}_t: t = 1,..., t_{max}\}$ using NMF.
Afterwards, we have a sequence of node-by-role matrices $\{\mathbf{G}_1,\mathbf{G}_{2},...,\mathbf{G}_{t_{max}}\}$ where each active node at time $t$ is represented with their current role memberships. The structural roles provide an intuitive representation for nodes that is scalable and efficient to compute for dynamic networks.

\subsection{Network Dynamics}\label{sec:role-dynamics}

Network dynamics refers to the structural patterns present in the network as it evolves over time. We posit that a subset of the learned structural patterns (behavioral roles) will become more or less important over time. Intuitively, a role may be active in a dynamic network only up to time $t_k$, at this point in time, the role might become inactive, and a new role may emerge or a role that is currently active may become more important (as the probability mass from the previous role is shifted to the current set of active roles). 
However, this process can only occur if we have a representative set of features over the entire time period. 
Suppose we extract features and learn roles from the first few timesteps, then in the future, there could be novel or more complex structural patterns that have not been represented. In that sense, we are interested in analyzing whether the behavioral roles represent basic generalizable patterns such as a bridge node, peripheral node, or the center of a $k$-star, or if they represent more complex patterns that are prone to drift as the network evolves.

The global network dynamics are analyzed in two ways. First, we analyze the role and network dynamics using the previous formulation. The idea is that over time the dynamic roles may drift; the role probability mass over the entire network shifts over time.  The second way is by learning a single global set of roles, then tracking these roles as they become more active (or important) or less active over time. One might expect that certain roles would appear and disappear over time. The second method for learning a single set of global roles is formally defined below.

\vspace{2mm}\noindent 
\textit{Global Features.} For each graph $\mA_t$, we extract a set of features denoted $\mathcal{L}_t$. 
The result is a sequence of feature lists $\{\mathcal{L}_1, \mathcal{L}_2, ..., \mathcal{L}_{t_{max}}\}$, then we take the union of the feature sets $\mathcal{L}^{\star} = \mathcal{L}_1 \bigcup \mathcal{L}_2 \bigcup ... \bigcup \mathcal{L}_{t_{max}}$ giving us the set of unique features over time.
Using the list of unique features $\mathcal{L}^{\star}$, we extract these features from each network snapshot resulting in a sequence of node-by-feature matrices $\mathcal{V} = \{\mathbf{V}_t: t = 1,..., t_{max}\}$ such that each $\mathbf{V}_t \in \mathbb{R}^{n \times f}$.

\vspace{2mm}\noindent 
\textit{Global Behavioral Roles.} Using the sequence of node-by-feature matrices, we construct a single global node-by-feature matrix $\mathbf{V}_g \in \mathbb{R}^{(n \times t_{max}) \times f}$ by stacking the node-by-feature matrices $\{\mathbf{V}_1, \mathbf{V}_2,...,\mathbf{V}_{t_{max}}\}$. 
We factor $V_g$ to discover $F_g$ and use this matrix to iteratively estimate the node-role memberships $\mathcal{G} = \{\mathbf{G}_t: t = 1,...,t_{max}\}$ given $F_g$ and $\mathbf{V} = \{\mathbf{V}_t: t = 1,..., t_{max}\}$ using NMF. Afterwards, we have a sequence of node-role matrices $\mathcal{G}$ where each active node at time $t$ is represented with their current role memberships.

\vspace{2mm}\noindent 
\textit{Role Importance.} Intuitively, a role may be useful for some subset of consecutive timesteps $t_j,...,t_k$, but then may become inactive and nodes that were previously assigned to the role may take on another more appropriate role. For this purpose, we define the relative role importance of the set of roles at time $t$ as,
\[\vx_t = \mG_t^{T} \ones \big/ n_t\]

\noindent where $\ones$ is a vector of ones and $\mG_t^{T}$ is the transposed of the node-by-role matrix at time $t$. The result is a sum of probabilities for each role over all active nodes. 
Naturally, if a role $i$ goes inactive or becomes stale, then this roles importance decreases, whereas if a lot of nodes actively take on role $i$ then this roles importance increases.

\subsection{Node Dynamics}
Node dynamics refers to the evolution of structural patterns for individual nodes. We posit that the nodes structural behavioral dynamics are non-stationary, that is they change or fluctuate over time. Of course, in social networks, a nodes behavioral dynamics may drastically change over a few years whereas in other types of communication networks a nodes behavior may stay relatively stable over time. 
As an example, the structure induced by emails for a given user may change during the work hours. Perhaps this user serves as a coordinator at work and therefore during the day their email activity represents structural behaviors such as the center of a star (node with large number of incoming or outgoing edges) or a bridge that connects multiple communities (or departments in this case).
In the case of biological networks, a nodes behavioral patterns may consistently oscillate or fluctuate over time, but the underlying behavior may not drastically change.

For tracking the structural patterns of individual nodes, we use the previous methods to analyze the importance of roles over time. Using this notion, we can naturally observe when a node has increasing or decreasing trends of structural behavior (e.g., becomes more social over time), as well as periodicity (e.g., takes on certain roles during the weekdays versus the weekends or at work versus at home), or if the node dynamics are relatively stable.

Besides tracking node dynamics, one might want to detect if the dynamics of an individual node change and the time at which this change occurred.
A simple approach might consider the similarity between the role membership vector for a specific node across time. For instance, if the node's role membership vector at time $t$ is different from their previous role membership vector at time $t-1$, then this indicates that the node has changed behavior. We briefly discuss this problem of detecting node anomalies in $\S$\ref{sec:discussion}.  

\begin{table}[b!]
\vspace{-2mm}
\caption{Performance of our role-based dynamic network analysis approach versus the state-of-the-art dynamic Mixed-Membership Stochastic Blockmodel (dMMSB). The dMMSB takes a day to handle 1,000 nodes~\cite{xing2010state}, while our model takes only 8.44 minutes for 8,000 nodes.}
\vspace{-2mm}
\label{table:scalability}
\begin{center}
\begin{small}
\begin{tabular}{  l  r  r  r }
\toprule 
\multicolumn{1}{c}{\TT \BB  \textbf{Dataset}} & \textbf{Nodes} & \textbf{Edges} & \textbf{Performance} \\
\midrule
\TT \BB \multirow{1}{*}{\rotatebox{0}{\textsc{Twitter}}} &  8,581 & 27,889 &  506.61 seconds\\ 
\TT \BB \multirow{1}{*}{\rotatebox{0}{\textsc{Network-Trace}}} & 183,389 & 1,631,824 & 16,138.71 seconds\\ 
\bottomrule
\end{tabular}
\end{small}
\end{center}
      \vspace{-4mm}
\end{table}

\subsection{Scalability and Practical Issues}
The structural dynamics approach is linear in the number of edges. The complexity can be stated as $O(|E| \cdot |T|)$ where $|T|$ is some trivial factor (even in the case where we use minute timesteps for analyzing IP-traces). A more accurate upperbound on the complexity can be defined in terms of the maximum number of edges at any given timestep. In this case, we can state the complexity as, 
\[O\Big(\max_t (|E|_t) \cdot |T|\Big)\]

The structural dynamics approach can handle very large networks consisting of millions of nodes and edges. The method can be used in practice for analyzing many very large real-world networks such as social networks, communication networks, citation networks, among many others.
This is in contrast to other recently proposed techniques such as the dMMSB~\cite{xing2010state,fu2009dynamic}. These models are quadratic in the number of nodes and thus unable to handle large networks. These models have been typically investigated on trivial sized networks of 18 nodes up to 1,000 nodes. 
Therefore, these models are unable to scale to the realistic networks with the number of nodes and edges in the millions.

Moreover, the dMMSB can handle 1,000 nodes in a day \cite{xing2010state} (See page 30), while our exploratory analysis approach for dynamic networks handles $\approx$8,000 nodes in 506.61 seconds (or 8 minutes and 26 seconds) shown in Table~\ref{table:scalability}. We provide performance results for other larger datasets of up to 183,389 nodes and 1,631,824 edges. In all cases, even for these large networks with over a million edges, our approach takes less than a day to compute and the performance results show the linearity of our method in the number of edges. For recording the performance results, we applied our method using a commodity machine Intel Core i7 @2.7Ghz with 8Gb of memory.

The proposed framework is also trivially parallelizable as features and behavioral roles can be learned independently at each timestep. This parallelization makes our method even more attractive and applicable for real-time analysis of the trends and patterns of communication and social networks. Furthermore, the proposed framework can naturally be applied in a streaming fashion. The role definitions can be adaptively updated in a streaming fashion by monitoring the error.

\section{Exploratory Analysis}\label{sec:results}
We demonstrate the utility of our approach for tracking the dynamics of the network as a whole ($\S$\ref{sec:network-dynamics}) and the dynamics of individual nodes ($\S$\ref{sec:node-dynamics}). 
The approach lends itself to visualizations that clearly show the local node and global network dynamics.
Our results clearly indicate that the behavior of nodes and the entire network as a whole are changing over time, i.e., non-stationary.
More specifically, we uncover the important dynamic patterns present in each of the communication, technological, and social networks.
Overall, we find that the node and network dynamics in each domain are quite different. 
The remainder of this paper explores these differences in dynamics, among the other more specific behavioral questions posed below.

In particular, for each type of network (social, biological, or technological), we seek to answer a few of the following questions.
What are the characteristics of nodes with respect to their learned behavioral roles? 
Does a node change slowly over time (days, weeks) or do nodes change behaviors continually throughout the day? 
and are the behavioral role changes predictable? 
Is there a normal progression of roles and are they cyclical (e.g., a role is exhibited in the morning, another in the afternoon, ...)? 
Is the behavior of nodes very stable over time or does it change a lot? 
What patterns of behavior are there? 
and are there local and global trends in the evolving behaviors? 
Are roles generalizable across time or do roles drift over time?

\begin{figure}[t!]
\centering
\vspace{-0mm}
\hspace{-7mm}
\includegraphics[width=3.5in]{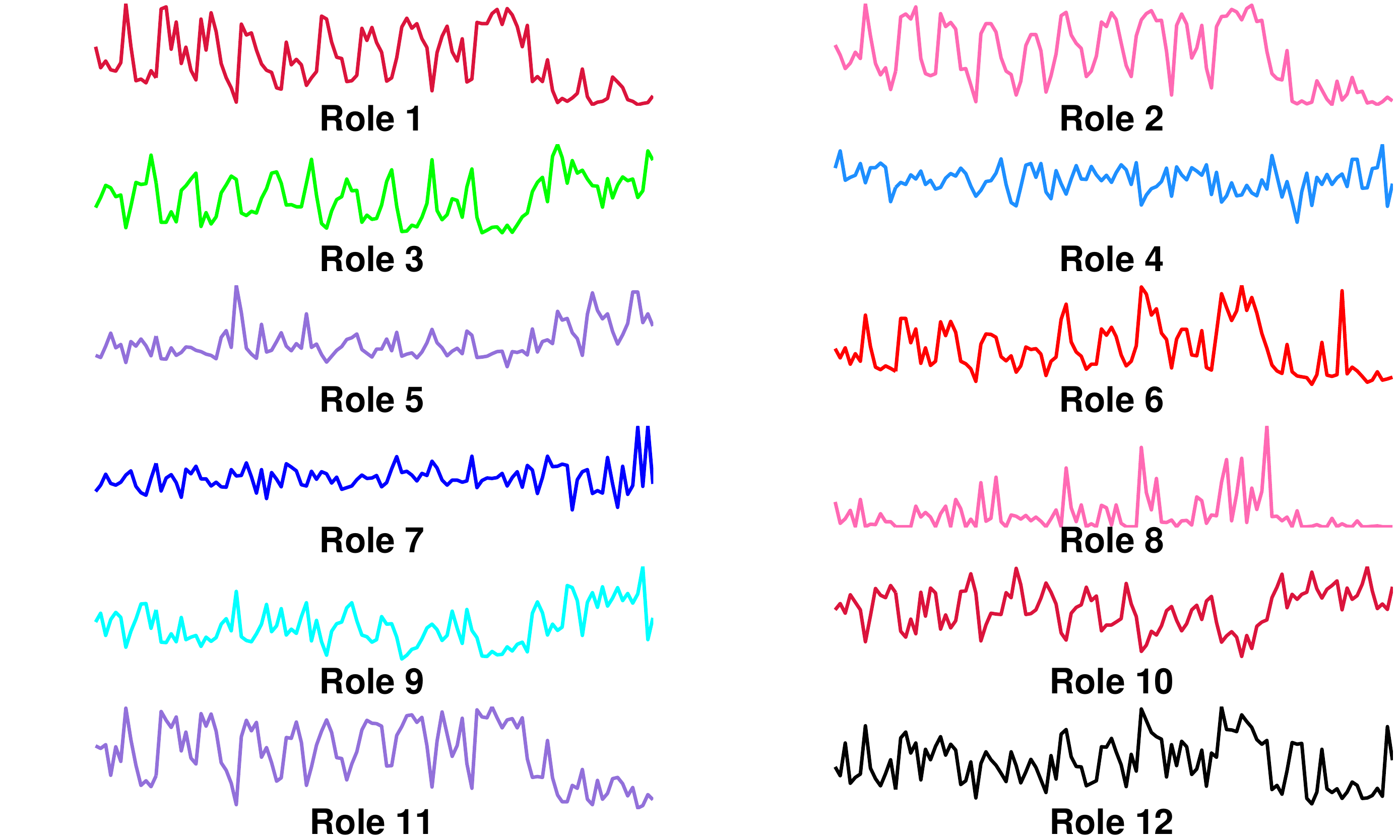}
   \caption{\textbf{Global Network Dynamics} of the Twitter ``Copenhagen" social network. The x-axis is time and the y-axis is the relative role importance.}
  \label{fig:twitter-role-dynamics}
  \vspace{-0mm}
\end{figure}

\begin{figure*}[t!]
\centering
\vspace{0mm}
\subfigure
    	{\label{fig:ip-trace-nodesense}\includegraphics[width=5in]{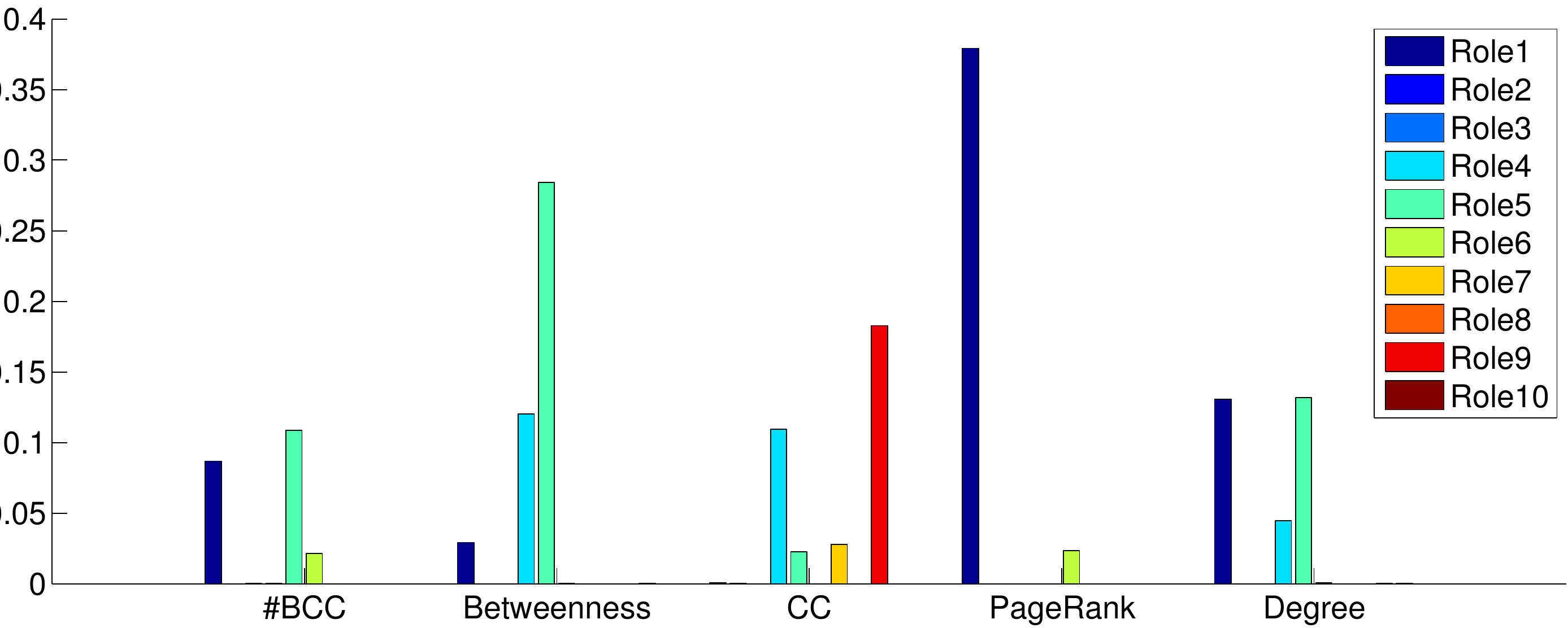}} 
\subfigure
    	{\label{fig:iptrace-mm}} 
    	\vspace{-3mm}
\subfigure
    	{\label{fig:iptrace-mm1}\includegraphics[width=6.15in]{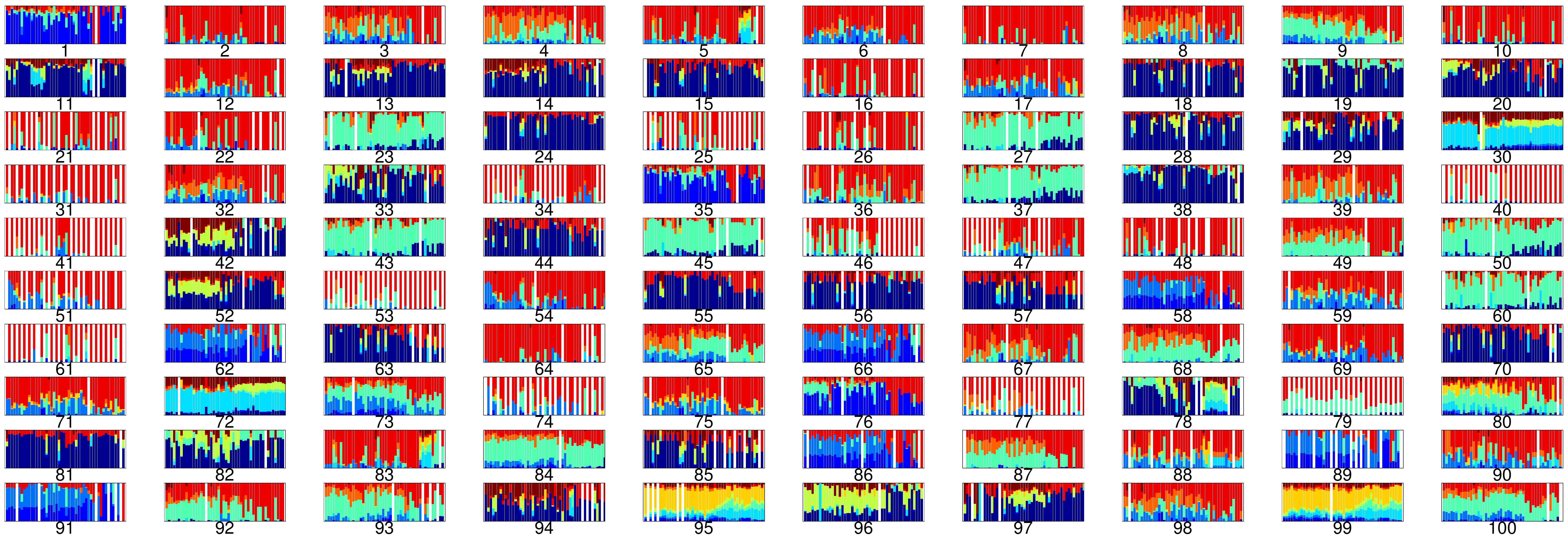}} 
\vspace{-3mm}
\subfigure
    	{\label{fig:iptrace-mm2}\includegraphics[width=6.15in]{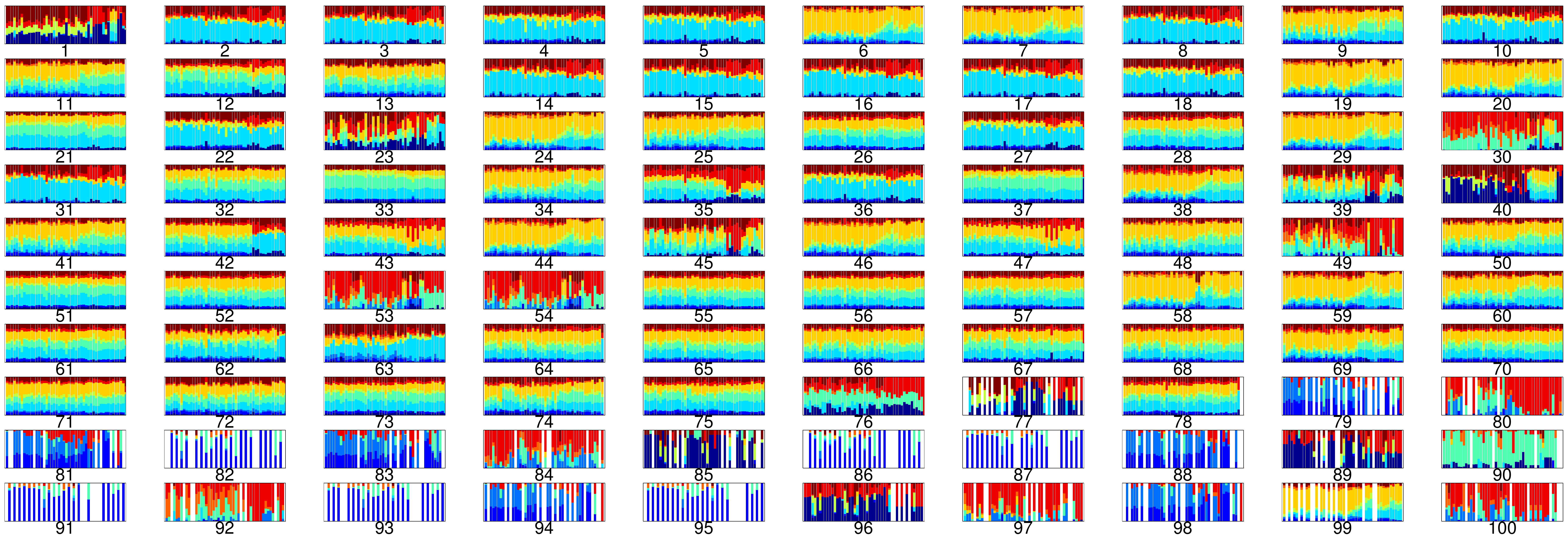}} 
\vspace{-3mm}
\subfigure
    	{\label{fig:iptrace-mm3}\includegraphics[width=6.15in]{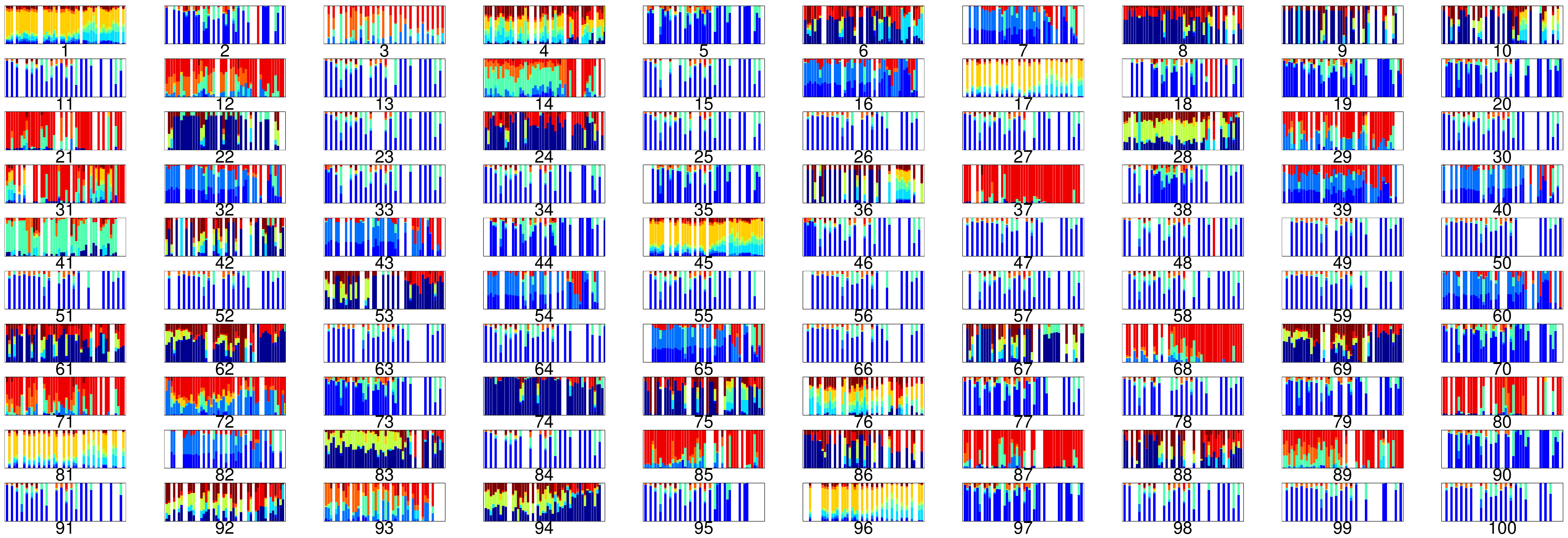}} 
    	\vspace{-0mm}
\caption{\textbf{Evolution of individual nodes.} The structural dynamics framework allows us to uncover important patterns of behavior in a large IP communications network. 
The roles are interpreted with respect to traditional structural properties and the role dynamics of 300 nodes are visualized where each color represents a specific behavioral role.
The x-axis is time and the y-axis is the mixed-memberships.
}
\label{fig:iptrace-mm}
\vspace{-5mm}
\end{figure*}

\subsection{Datasets}
The first two datasets are used to analyze the structural behavior of nodes over time while the last dataset is used to analyze the global network dynamics.

\vspace{2mm}
\noindent\textbf{Email (University).}
This network data consists of university emails from two weeks~\cite{ahmed-TR}. The email network (who-emails-whom) was generated using email logs from the university mailservers. We only consider email accounts that have at least one incoming or outgoing edge in the trace. We used a subset of this network consisting of 116,893 nodes and 1,270,285 edges across 50 timesteps where each timestep represents 1 hour of activity. From this network, 652 features were automatically extracted from the initial timestep which resulted in learning 10 behavioral roles.

\vspace{1mm}
\noindent\textbf{Enterprise Network Traces.}
We use real network-trace data sets collected over time on an enterprise network. The nodes are IP addresses and links are communications between IPs. Each communication has a begin-time and an end-time. The resulting network used for analysis consists of 183,389 nodes that have 1,270,285 edges between them over 49 timesteps of 15 minutes a piece. From this network, 268 features were automatically extracted from the initial timestep which resulted in learning 11 behavioral roles.

\vspace{1mm}
\noindent\textbf{Twitter (Copenhagen).} The network is formed by the set of 74,227 reply-to-messages in the \#cop15 Twitter hashtag occurring over a two-week period from 12/07/09-12/18/09. See~\cite{ahmed2010reconsidering,ahmed2010time} for more details.
Using a subset of this data, we constructed a network of 8,581 nodes consisting of 27,889 edges from 112 timesteps of 3 hours a piece. From the entire network, 729 features were automatically extracted, from which 12 behavioral roles were learned over time.


\subsection{Global Network Dynamics} \label{sec:network-dynamics}
We investigate the collective network dynamics of the Twitter social network (See $\S$\ref{sec:role-dynamics} for algorithmic details) formed from the two week United Nations Climate Change Conference in 2009.
Figure~\ref{fig:twitter-role-dynamics} plots the relative importance of each role over time. 
This visualization clearly shows the changes in the global network dynamics.
From this analysis, we find that the relative importance of some of the roles in Twitter naturally fluctuate between night and day (see roles 1 \& 2), while other roles are more stable or stationary (roles 5 \& 7), and therefore can be easily predicted.
However, we also find roles that are more volatile, with seemingly no regularity. 
Role 8 is a prime example. 
This role represents a more complex connectivity pattern that arises infrequently.
Thus, when the importance of this role spikes, it could indicate the presence of a global network anomaly.

A spike in the importance of role 6 is shown towards the last timesteps. 
This pattern is relatively different from the other smaller or more common cyclical spikes in the other roles. 
We also observe a type of \textit{step pattern} in the importance of roles 1 and 2 (among others). 
This step pattern is located towards the last few timesteps. 
Moreover, there is also upward step patterns (see the importance of role 10). 

Perhaps more importantly, towards the end of time, there is a relatively common increasing and decreasing trend in the role importances. 
The increasing trend is most strikingly seen in role 9 or 10 whereas the decreasing trend is most strikingly seen in role 1 and 2.
Interestingly, we examined the twitter action log from this period of time and found that the communication patterns between the twitter users involved in the climate change conference drastically changed. 
Since the conference was coming to an end, there were less users tweeting about the cop15 UN climate change conference, and therefore the users that were still actively tweeting became more personal with one another, forming more densely connected subgraphs. 
Nevertheless, as shown, our approach captures these dynamical patterns that agree with human intuition in a fast completely automated manner.

\begin{figure}[t!]
\centering
\hspace{-2mm}\subfigure[Time-evolving Mixed-Memberships (Email) \;\;\;\;\;\;\;\;\;\;\;\;\;\;\;\;]
    	{\label{fig:email-mm}\includegraphics[width=4in, bb= 100    65   850   710, clip=true]{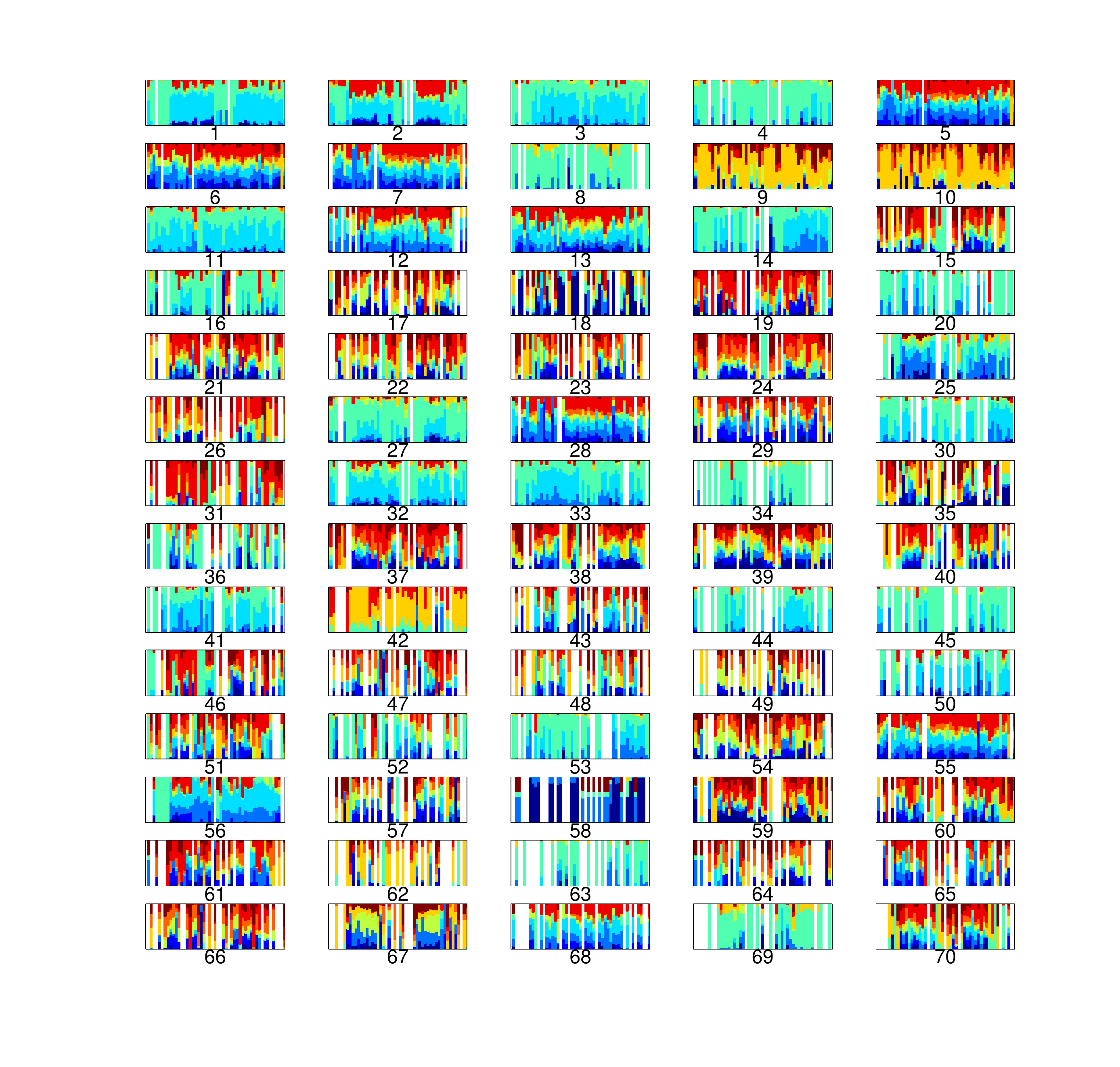}}
\hspace{-0mm}\subfigure[Email Role Interpretation]
    	{\label{fig:email-role-interp}\hspace{-3mm}\includegraphics[width=3.4in]{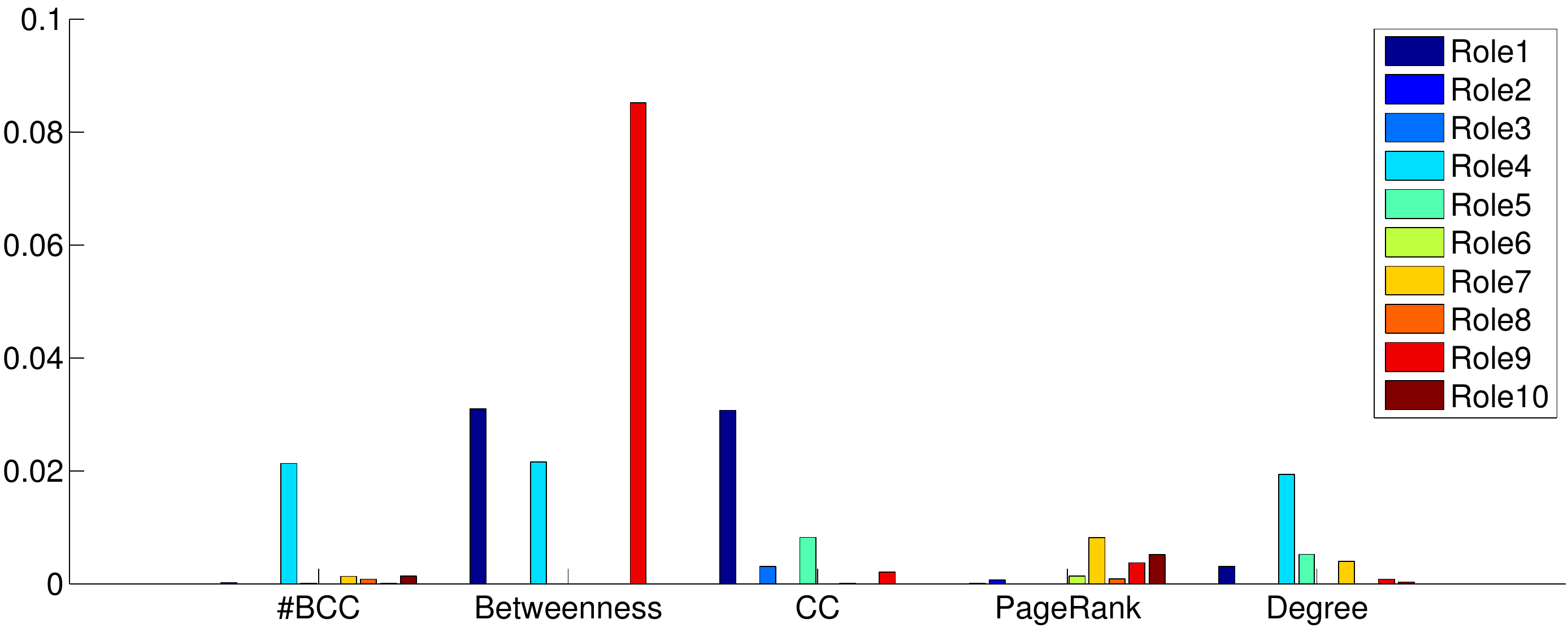}}
    	\vspace{-2mm}
\caption{The structural dynamics framework allows us to uncover important patterns of behavior in an email network. 
(a) visualizes the structural dynamics of individual nodes over time where each color represents a role. The x-axis is time and the y-axis is the proportion of each structural behaviors. (b) is the interpretation of the structural patterns with respect to traditional structural properties such as betweenness, biconnected components, pagerank, clustering coefficient, and degree.}
\label{fig:email}
\vspace{-0mm}
\end{figure}

\subsection{Local Node Dynamics} \label{sec:node-dynamics}
We systematically analyze the evolutionary behavioral patterns of individual nodes in two large real-world networks, namely, an email communication and an IP-to-IP network.
Given a time-series of role-memberships for an individual node, we can identify unique roles for that node and detect when a node's behavior deviates from their past behaviors. 
Additionally, we interpret the roles with respect to traditional node measures~\cite{rolX2011-TR}.
Formally, the roles are interpreted using the dynamic node-by-role memberships $\mG_t$ and a node measure matrix $\mM_t \in \mathbb{R}^{n_t \times m}$ to compute a non-negative matrix $\mE_t$ such that $\mG_t \mE_t \approx \mM_t$ . The node measurements used are betweenness, biconnected components, pagerank, clustering coefficient, and degree. The matrix $\mE_t$ represents the contributions of the traditional node measures to the roles at time $t$. The contributions are averaged across time.

The evolving behavioral patterns for a large set of nodes in the IP-trace network and the interpretations of their dynamic roles are shown in Figure~\ref{fig:iptrace-mm}. 
From the role interpretation, we most clearly see that the eighth role represents high clustering, while the fifth role represents betweenness, whereas the first role represents mainly nodes with high pagerank.
The other roles represent more specialized structural motifs as they are represented by a combination of structural characteristics.

Strikingly, we find approximately four major evolutionary patterns for the individual nodes. 
For instance, there are nodes whose structural patterns are relatively stationary over time. 
In particular, we see nodes that are consistently the red role, which can be interpreted as nodes with high clustering coefficient (and other similar structural properties).
We also find nodes that consists of mostly role 4 and role 9 which represent betweenness and clustering coefficient, respectively. 
Moreover, there is a slight downward trend in the structural pattern representing a combination of betweenness and clustering coefficient (role 4).
There are also many other interesting patterns such as spikes in certain roles, cycles, and upward and downward trends in the structural behaviors of nodes over time. 
Additionally, we also find nodes that contain interesting patterns with respect to their activity and inactivity. 
In particular, nodes become active and then inactive frequently (inactivity is represented as white).

Indeed, the structural dynamics approach can be used to understand the evolution of many real-world networks. 
Figure~\ref{fig:email} visualizes the node dynamics for the email communication network and interprets the corresponding structural patterns.
Just as before, we can identify significant trends and patterns and interpret these using the role interpretations from Figure~\ref{fig:email-role-interp}.
For instance, node 67 has a sequence of stable roles over two time periods. 
In between these two time periods, there is inactivity. 
This is not surprising as the evolving mixed-memberships represent only two days of email communications. 
The sequence of stable mixed roles for node 67 represent the email activity during the daytime, whereas the inactivity in between these two periods of time represents the night.
Similar patterns can be seen in other nodes.
We can also identify nodes that transition to a different set of roles or take on more of these roles at night, such as 1, 2, 39, and many others. 
We also find nodes that have inconsistent behavior over the time, such as nodes 17, 18, and 19, among others. 
The nodes with inconsistent behavior could indicate anomalous activity. 
Furthermore, we also find nodes that have relatively stable structural behavior over the two days, such as nodes 5 and 6. 
This type of behavior is also unusual (since we would expect a nodes roles to transition from the daytime work hours to nighttime).
However, nodes that are consistently dominated by multiple active roles are of importance, since they connect to groups of nodes with different types of connectivity patterns (see nodes 5-7).
Indicating that these individuals may serve in managerial or leadership roles.

Our approach clearly shows how the behavior of local nodes change over time.
We also identify the differences in the dynamical patterns of these two networks.
For instance, the individual node dynamics in the email communication network are more volatile than the IP-trace network. 
Moreover, the individual nodes in the IP-trace network take on less roles, while the nodes in the email network are often dominated by multiple active roles.
Nevertheless, our method also captures similarities between the individual nodes in the two networks. 
Since the dynamics of nodes in both networks are controlled by humans, they naturally share similar behavioral fluctuations (between night/day or weekday/weekend).

\section{Related Work}\label{sec:related-work}
While there is a lot of work on dynamic graph patterns~\cite{götz2009modeling,leskovec2005graphs,leskovec2007dynamics,papadimitriou2005streaming,sun2007graphscope}, temporal link prediction~\cite{dunlavy2011temporal}, anomaly detection~\cite{abello2010detecting},  dynamic communities~\cite{lin2009analyzing,tang2008community,greene2010tracking}, and many others~\cite{yang2011patterns,habiba2008finding,o2005prediction}.
No one has yet to propose a scalable role-based analysis framework for large time-varying networks.
The closest work is that of \cite{fu2009dynamic,xing2010state} where they develop the dMMSB model (based on a completely different process) for small graphs. Their model is capable of handling 1,000 nodes in approximately 1 day while our approach is linear in the number of edges and capable of handling 1,000 nodes in only a few minutes (practical for large real-world networks).

\section{Discussion}\label{sec:discussion}
A network measure captures a particular feature of the network topology.
For instance, social networks have a large clustering coefficient~\cite{watts1998collective} while biological and technological networks have been found to have negative assortativity~\cite{newmanassort:2002}.
However, these features tell us only about a single pattern present in these networks, missing perhaps more important patterns of connectivity.
Moreover, as societies change and biological systems evolve, these properties may no longer be of importance.
Naturally, these dynamic networks induce arbitrary patterns of connectivity that are challenging, but important to identify.
The traditional network measures usually capture simple connectivity patterns that can be understood quickly, while more complicated, less frequent patterns might actually be of more importance.

Instead of a fully-automatic approach for mining large dynamic networks (like the one proposed in this paper), one might imagine simply selecting features that are important for each type of network and tracking these over time.
However, simply selecting the important properties for each type of network is incredibly difficult as the connectivity patterns and properties of these networks are not fully understood, and moreover, the network measures only capture simple connectivity patterns, making this manual selection impossible.
Furthermore, this task requires expert knowledge in that domain (biological, technological, or social networks), and there is no guarantee that these features are representative of the important structures that are truly changing over time.
The properties also must be fast to compute (linear in the number of nodes or edges) and representative of the important connectivity patterns, which is also challenging.

The main advantage of using a manually tuned simplistic approach over \textsc{Role-Dynamics} is interpretability, while the disadvantages of such an approach are extensive. For instance, manually selecting measures for a specific network would be costly in time/money, inaccurate, possibly slow, it would not be adaptive, or able to capture novel connectivity patterns over time.
Furthermore, in real-time networks where the connectivity patterns are non-stationary and changing very quickly, then even having an expert who can somehow select the important features is not enough as they would change before the system could be retuned.
The algorithm proposed in this paper is applicable for any type network, completely automatic (requiring no user-defined parameter), fast for large dynamic networks, while capturing arbitrary connectivity patterns that are important for the given network.
The main disadvantage of our approach is in the interpretation of the patterns over time.
However, in this paper, we have used analytical tools to interpret and understand the dynamic patterns with respect to more traditional measures that have been widely studied by many researchers.
In future work, we plan to build more sophisticated tools for analyzing and visualizing the important connectivity patterns present in each type of network.

Nevertheless, the \textsc{Role-Dynamics} approach is a prime candidate for other applications such as real-time graph-based anomaly detection~\cite{noble2003graph}, dynamic relational classification~\cite{neville:tenc}, and for predicting future structural patterns.
The goal of anomaly detection in graphs is to detect nodes, links, or network states that are anomalous, and therefore the actual interpretation of the learned patterns from \textsc{Role-Dynamics} are no longer important (or of secondary importance for forensics).
The majority of graph-based anomaly detection methods define a set of particular connectivity patterns such as degree or clustering coefficient~\cite{rattAnomKDDExpl05,akoglu2010oddball,noble2003graph,ide2004eigenspace}. 
Therefore, these methods only capture specific known network anomalies, but fail to capture novel anomalies. 

However, our proposed approach for exploring dynamic networks is non-parametric and therefore learns a representative set of graph features that generalize over a class of time-evolving networks. Thus, it is capable of capturing novel anomalies, making it suitable for IP communication networks where attack-vectors may be novel and therefore unknown to even specialists. 
We plan to use \textsc{Role-Dynamics} for detecting \textit{node anomalies} (e.g., if a specific node takes on unusual roles) and \textit{network anomalies} (e.g., if the behavior of the entire network changes) in future work.

\section{Conclusion}\label{sec:conclusion}
We proposed an efficient and scalable framework for mining the structural dynamics of large real-world networks. 
Our novel approach tracks changes in local (node-level) and global behavior over time.
In particular, our approach captures arbitrary patterns of connectivity, requires no user-defined parameters, interpretable, and is fast for mining large networks (linear in the number of edges).
The approach lends itself to visualizations that clearly show how the behavior of the network as a whole (Fig.~\ref{fig:twitter-role-dynamics}) and the individual nodes (Fig.~\ref{fig:iptrace-mm} \& \ref{fig:email}) change over time.
The proposed \textsc{Role-Dynamics} framework can be used as a basis for more sophisticated models and analysis tools.

In future work, we plan to model the behavioral transitions over time and apply this model for prediction tasks like classification, anomaly detection, and for predicting future structure. We also plan to develop a clustering algorithm around this framework to group nodes based on their learned structural dynamics. Modeling the dynamics and the transition patterns of individual nodes will increase the effectiveness of our exploratory analysis framework.

\section*{Acknowledgments}
This work was performed under the auspices of the U.S. Department of Energy by Lawrence Livermore National Laboratory under contract No. DE-AC52-07NA27344. 
This research is also supported by NSF under contract numbers IIS-0916686, IIS-1017898, and SES-0823313. This research was also made with Government support under and awarded by DoD, Air Force Office of Scientific Research, National Defense Science and Engineering Graduate (NDSEG) Fellowship, 32 CFR 168a. The U.S. Government is authorized to reproduce and distribute reprints for governmental purposes notwithstanding any copyright notation hereon. The views and conclusions contained herein are those of the authors and should not be interpreted as necessarily representing the official policies or endorsements either expressed or implied, of LLNL, DOE, NSF, or the U.S. Government.

\bibliographystyle{abbrv}
\bibliography{rossi}

\end{document}